\documentstyle[prl,aps,preprint]{revtex}
\def\thru#1{\mathrel{\mathop{#1\!\!\!/}}}
\begin{document}
\tighten

\title{WILSON'S EXPANSION WITH POWER ACCURACY\thanks
{This work is supported in part by funds provided by the U.S.
Department of Energy (D.O.E.) under cooperative agreement
\#DF-FC02-94ER40818.}}

\author{Xiangdong Ji}

\address{Center for Theoretical Physics \\
Laboratory for Nuclear Science \\
and Department of Physics \\
Massachusetts Institute of Technology \\
Cambridge, Massachusetts 02139 \\
{~}}

\date{MIT-CTP-2437 \hskip 1in hep-ph/xxxxxxx \hskip 1in June 1995}

\maketitle

\begin{abstract}
Because of the infrared renormalons, it is difficult
to get power accuracy in the traditional approach to
the Wilson's operator product expansion. Based on
a new perturbative renormalization scheme for
power-divergent operators, I propose
a practical version of the OPE that allows
to calculate power corrections to desired accuracy.
The method is applied to the expansion of
the vector-current correlation function in QCD vacuum,
in which field theoretical status of the gluon
condensate is discussed.

\end{abstract}

\pacs{xxxxxx}

\narrowtext
Wilson's operator product expansion (OPE) \cite{WIL}
has become
one of the most powerful tools in modern field theory.
Indeed, many graduate text books on field theory
devote an entire chapter treating the subject \cite{MUTA}.
Direct applications of the OPE in Quantum Chromodynamics
(QCD) include deep-inelastic scattering,
heavy-quark expansion, weak-nonleptonic decays,
and QCD sum rule calculations, etc. Generally speaking,
the OPE provides the starting-point for any perturbative
calculations in hard processes. The underlying principle of
the OPE---scale separation and factorization---
has spawned many uses of the so-called
effective field theories in nuclear and particle physics.

Given the success of the OPE, it may be surprising to
realize that the standard text-book approach to the expansion
is too formal to be used in some applications. A problem
occurs when one seriously considers the so-called {\it power
corrections} to the leading term in the expansion.
To explain, let me first remind the reader some
basics of Wilson's expansion \cite{MUTA}.
Consider a time-ordered product of two currents
separated by, say, an Euclidean short distance $\xi^2$.
According to Wilson, the product can be expanded as,
\begin{equation}
     T[J(\xi)J(0)]
     = \sum_{i=0}^{\infty} C_i(\xi^2,\mu^2) O_i(\mu^2)  \ ,
\end{equation}
where, for simplicity, I have neglected possible Lorentz indices.
The short distance physics above the factorization scale $\mu^2$
is included in the coefficient functions $C_i$, ordered
according to the descending singularities as $\xi^2\to 0$.
$O_i(\mu^2)$ are renormalized operators with increasing dimensions,
designed to account for long distance physics below
$\mu^2$. The $\mu^2$ dependences are cancelled
between $C_i$ and $O_i$ term by term in the expansion.

The expansion in principle can serve to define the composite
operators $O_i$ if one knows how to classify the short
distance singularities of an operator product. In practice,
however, a reverse procedure is followed in constructing
the expansion:
A tower of operators with right quantum numbers
are first chosen, and then the coefficient functions
are {\it calculated} by sandwiching the expansion in
a set of perturbative states. Implicitly assumed
in this standard procedure, though not essential to
the principle of the OPE,
is that the power divergences in the operators $O_i$
are subtracted with a {\it perturbative normal ordering}.
For the operators with vacuum quantum numbers, such as $F^2=
F^{\alpha\beta}F_{\alpha\beta}$, it
means the subtraction of the perturbative-vacuum expectation value.
 For others, it means use of the dimensional
regularization in which
the power divergent integrals $\int d^dk/k^\alpha=0$
are taken to be zero.

It was first pointed out by t' Hooft
\cite{HOO} that the coefficient functions thus
obtained,
$C_i(\alpha_s) =\sum_n \alpha_s^n c_{in}$,
may contain the {\it infrared (IR) renormalons}---a jargon for
$c_{in} \sim n!$ to grow factorially with a fixed sign.
Although this has never been proved conclusively in QCD except
in certain limiting cases such as the number of flavors
$N_f\to \infty$, many have taken the observation
seriously. If true, $C_i$ is a non-Borel summable
series and is genuinely infinite.
If treating $C_i$ as an asymptotic series,
it is easy to see the uncertainty in regularizing
the series, such as truncating it around the minimal
term, or regulating the singularity in the Borel plane after
a Borel transformation,
spoils a clear-cut classification of $C_i(\xi^2)$
in their singularities. At practical
level, the power corrections to the OPE become
ill-defined due to the uncertainty in the leading-order
coefficient function.

For applications where power corrections are small,
like heavy-quark expansion for the top flavor,
the problem is of only formal importance.
However, there are applications where the understanding
of power corrections is absolutely essential.
In extracting the strong coupling constant $\alpha_s$
from the deep-inelastic sum rules, such as Bjorken
and Gross-Llewellyn Smith sum rules, the so-called
higher-twist contributions are an important source
of theoretical uncertainty \cite{JI1}.
In the charm and bottom systems, $1/m_Q$ corrections
to Isgure-Wise symmetry relations \cite{IW}
valid in the heavy quark limit are substantial.
Finally, the entire QCD sum rule phenomenology
initiated by Shifman, Vainshtein, and Zakharov \cite{SVZ}
relies on a meaningful understanding of power corrections
at fundamental level.

The IR renormalons in QCD are generated from perturbative
calculations in
infrared regions of Feynman diagrams. An indication
for this is that the high-order diagrams
effectively renormalize the coupling constant appearing
in the lower-orders. The infrared
Landau singularity in the resulting running coupling
is essentially the source for the $n!$ behaviors
at high orders \cite{ZAK}. In light of this observation,
the appearance of the IR renormalons in the OPE is
somewhat surprising, for the OPE is devised precisely to account for
soft regions of perturbative
diagrams with the matrix elements of composite operators.
The puzzle was solved first by F. David \cite{DAV}, who showed that the
power-divergent composite operators
defined with the standard subtraction scheme
also contain the IR renormalons.
And the IR renormalon singularities
in the coefficient functions and operators matrix elements
cancel.

Although the IR renormalons do not invalidate the
standard form of the OPE, they pose
practical problems in carrying out the expansion
to power accuracy. The coefficient functions
and the power-divergent composite operators
are both ill-defined, and one must
find a consistent way to regularize them.
Grunberg \cite{GRU} and Mueller \cite{MUL1} have
both suggested ways to regularize
the perturbation series in coefficient functions.
However, computations of higher-twist
contributions in their schemes appear difficult.
On the other hand, in a
recent work by Martinelli and Sachrajda\cite{MS},
a non-perturbative
renormalization of composite operators is proposed.
It remains to see, however, how practical the
method can be used to compute subtraction
in coefficient functions.

Since the IR renormalons arise from soft regions
in Feynman diagrams, a physical solution to the problem
is to avoid perturbative calculations in these regions.
The idea is not new, as it has been advocated by
Novikov et al. \cite{NOV} since many years ago.
However, the only serious
study in this direction was the work
by Mueller\cite{MUL2}, which has largely been ignored
in the literature. The main problems
with this approach, to my opinion, have been two:
first, how to make systematic soft subtraction
in the coefficient functions? second, how to
calculate the non-perturbative matrix elements of composite
operators consistent with the subtraction.
In this paper, I
hope to provide answers to both questions through
the example of the vector-current correlation function in
QCD vacuum.

To begin, let me recall once again the basic principle of the
OPE: the composite operators are introduced to account for
soft contributions in Feynman diagrams.
Thus if the OPE is applied in a perturbative state,
the composite operators shall contain perturbative
soft contributions. In the
standard regularization for the composite
operators, however, the perturbative
contributions are completely subtracted along
with the power divergences. This
is convenient for calculating the coefficient
functions, but the operators defined
in this way cannot account for
the perturbative soft contributions in
Feynman diagrams for correlation functions.
Therefore, a fundamental solution to
the IR renormalon problems is to find a practical way to
regularize power divergences of
composite operators while maintaining
their soft contributions to a perturbative calculation.

Composite operators can be defined entirely through their
insertions into Green's functions \cite{ZIM}. As such, the renormalization
of the Green's functions determines the renormalization
of the composite operators. According to the standard
renormalization theory, we need to consider
only the Green's functions with non-negative superficial
degree of divergence, which is defined
for a graph $G$ as, $
    \delta(G) = 4 - d_O - d_\phi n_\phi \ ,
$ where $d_O$ and $d_\phi$ are the dimensions
of the operator $O$ and field $\phi$, and
$n_\phi$ is the number of external $\phi$ lines
in the graph.
The counter term operators can be constructed
from the overall subtraction of the
primitively divergent Green's functions.
Now it is well-known that the renormalization subtraction
have large freedom with finite contributions (renormalization
group). In QCD, since the low momentum regions of a
Feynman diagram is not perturbatively calculable,
one shall not subtract divergences along with any
soft contributions. In the following, I call
such subtraction as {\it "minimal"
subtraction of power divergences.}

To illustrate a systematic way of doing minimal subtraction,
let me construct the
composite operator $[\phi^2]$ in the $\phi^4$ theory.
For our purpose, we are interested in only
one insertion of the operator. According to
power counting, two Green's functions
have superficial divergences,
$\langle 0|T\phi^2 \phi(x_1)\phi(x_2)|0\rangle$ and
$\langle 0|T\phi^2|0\rangle$, with superficial
degrees of divergence 0 and 2, respectively.
Therefore, the renormalized composite
operators can be defined as,
\begin{equation}
     [\phi^2] =  Z_2 \phi^2 + Z_0 I
\end{equation}
where $Z_i$ are renormalization factors.
$Z_2$ is determined by subtraction
of primitive log-divergent diagrams in
$\langle 0|T\phi^2 \phi(x_1)\phi(x_2)|0\rangle$.
On the other hand, $Z_0$ is determined by
primitive quadratically-divergent diagrams in $\langle 0|T\phi^2|0\rangle$.
In the standard approach, one writes, $
        Z_0 = - Z_2 \langle 0|T\phi^2|0\rangle_{\rm pert}
$. The operator $[\phi^2]$ defined
in this way depends on perturbative
calculations in soft regions. On the other hand,
The minimal subtraction can be defined as follows.
Consider any perturbative diagram for
$\langle 0|TZ_2\phi^2|0\rangle$.
Subtract only the contributions from regions
where all the momenta in the diagrams are larger than
some scale $\Lambda^2$. For instance, suppose
we have the following contribution to
$\langle 0 |TZ_2\phi^2|0\rangle$,
\begin{equation}
     \int^\infty_0 d^4k_1 \int^\infty_0 d^4k_2 ...
     \int^\infty_0 d^4k_n f(k_1,...,k_n)
\end{equation}
Then we subtract the part when all the integrations run from
$\Lambda^2$ to $\infty$,
\begin{equation}
     Z_0 = - \int^\infty_{\Lambda^2} \int^\infty_{\Lambda^2} ...
     \int^\infty_{\Lambda^2}
    d^4k_1d^4k_2...d^4k_n
   f(k_1,...,k_n)
\end{equation}
Thus, $Z_0$ contains only hard contributions.
It remains to see however, the remaining part of
the Green's functions, with mixtures of
soft and hard momentum flows, are finite.

The proof for the finite remaining part
is done by explicit construction.
It is easy to see through working with
two or higher-loop diagrams that
\begin{eqnarray}
      \langle 0|[\phi^2]|0\rangle_{\rm pert} && =
     \sum_{n=1}^{\infty} (-1)^{n+1} \int^{\mu^2}_0 d^4k_1...\int^{\mu^2}_0
d^4k_n
    \nonumber \\
     && \times \langle 0|TZ_2\phi^2 \phi(k_1)...\phi(k_n)\phi(-k_1-k_2...-k_n)
|0\rangle^{\rm trun} \nonumber \\
     && \times
   \langle 0|T \phi(k_1)...\phi(k_n)\phi(-k_1...-k_n)|0\rangle(\mu^2)
\label{soft}
\end{eqnarray}
where the first factor in the integrand
$\langle 0|TZ_2\phi^2 \phi(k_1)...\phi(k_n)\phi(-k_1...-k_n)|0\rangle^{\rm
trun}$ is a connected, truncated Green's
function with
external momenta $k_1$, $k_2$, ... ,$k_n$, $-k_1...-k_n$.
It is finite because all
logarithmic sub-divergences are cancelled by
the renormalization factor $Z_2$. The second factor
$\langle 0|T\phi(k_1)...\phi(k_n)\phi(-k_1...-k_n)
|0\rangle$ is an ordinary Green's function with all
internal lines restricted to $k^2>\mu^2$.

To illustrate the use of the minimal-subtracted operators
in the OPE,
I consider the vacuum correlation
function of the conserved vector currents in the
chiral limit ($m_q=0$)
$
     i \int e^{i\xi\cdot q}d^4\xi
         T[J_\mu(\xi)J_\nu(0)]
        = (-g_{\mu\nu}q^2 + q_\mu q_\nu)  \Pi(q^2) \ .
$
Define Adler's $D$-function through, $D(q^2) = -(2\pi)^2
q^2(d\Pi/dq^2)$. The OPE for $D(q^2)$ at the space-like $q^2$
reads ($Q^2=-q^2$),
\begin{equation}
D(Q^2)= C_0(\alpha_s, \Lambda^2))I
+ C_4(\alpha_s){2\pi\alpha_s
     F^2(\Lambda^2)\over 3Q^4} + ... \ ,
\label{exp}
\end{equation}
where $I$ is a unit operator and the renormalization scale $\mu^2$
is chosen to be $Q^2$. $F^2(\Lambda^2)$ is minimally subtracted
operator without the quartic divergence,
\begin{equation}
 F^2(\Lambda^2) = F^2_R - \langle 0|F^2_R|0\rangle_{\rm pert}
          ({\rm all} \ k^2> \Lambda^2) \ .
\end{equation}
The renormalized $F^2_R$ is free of
logarithmic divergences. In dimensional regularization
and with covariant gauge fixing,
we have \cite{JI20},
\begin{eqnarray}
    F^2_R = && (1 + g{\partial \ln Z_g \over \partial g}) F^2
         + (g{\partial \ln Z_g \over \partial g}
         - \lambda {\partial \ln Z_3 \over \partial \lambda}
         + {1\over 2}g{\partial \ln Z_3 \over \partial g})O_{GF} \nonumber \\
         &&+ (\lambda {\partial \ln \tilde Z \over \partial \lambda}
         - {1\over 2}g{\partial \ln \tilde Z \over \partial g})
         \partial^\mu \bar \omega D_\mu \omega
         + (\lambda {\partial \ln Z_2 \over \partial \lambda}
         - {1\over 2}g{\partial \ln  Z_2 \over \partial g})\bar \psi i\thru
D\psi
 \end{eqnarray}
where $\lambda$ is
a gauge parameter, and $Z_g$, $Z_2$, $Z_3$ and $\tilde Z$ are renormalization
constants for the gauge coupling $g$, quark fields $\psi$, gauge potential
$A^\mu$ and ghost field $\omega$, respectively. All operators on the
right-hand-side are unrenormalized and,
$      O_{GF} = -{1\over 2\lambda}(\partial^\mu A_\mu)^2
       - {1\over 2} {\delta S\over \delta A_\mu} A_\mu $,
where $S$ is the gauge-fixed QCD action.

Let us see how the coefficient function $C_0$ and the
non-perturbative matrix element $F^2(\Lambda^2)$
can be consistently calculated. First, sandwiching
the expansion in Eq. (\ref{exp})
in the perturbative vacuum state, we have,
\begin{equation}
      C_0(\alpha_s,\Lambda^2) = C_0(\alpha_s,0) -
       C_4(\alpha_s){2\pi\alpha_s\langle 0|F^2(\Lambda^2)
      |0\rangle_{\rm pert} \over 3Q^4}
\label{sub}
\end{equation}
where $C_0(\alpha_s,0)$ is the coefficient function
calculated when assuming the perturbative matrix element
of $F^2(\Lambda^2)$ vanish. The contributions to
$C_0(\alpha_s, 0)$ from one or two soft-gluon or ghost lines
are subtracted by the second term in Eq. (\ref{sub})
which, according to
our definition, can be computed through a formula like
in Eq. (\ref{soft}).

Since perturbative diagrams are most conveniently
evaluated in the dimensional regularization and modified
minimal subtraction scheme,
we shall study the OPE in this scheme (labelled by $\overline{\rm MS}$).
$C_0(\alpha_s, 0)$ has been calculated to
three-loops \cite{G} (assuming three flavors),
\begin{equation}
      C_0(\alpha_s, 0) = 1 + {\alpha_s\over \pi} + 1.6
  \left({\alpha_s\over \pi}\right)^2 + 6.4\left({\alpha_s\over \pi}\right)^3
 + ... \ ,
\end{equation}
and $C_4$ has been calculated at one-loop \cite{B},
$
      C_4(\alpha_s) = 1 + {7\alpha_s\over 6\pi} + ...
$. To illustrate the soft subtraction, I have
calculated the perturbative matrix element of $F^2(\Lambda^2)$
up to two-loops,
\begin{equation}
       \langle 0|F^2(\Lambda^2)|0\rangle_{\rm pert}^{\overline{\rm MS}}
      = {3\Lambda^4\over \pi^2}\left(1 + (2.41+1.25\ln{Q^2\over
     \Lambda^2}) {\alpha_s\over \pi} \right)
\end{equation}
Finally, we have coefficient function for the unit operator in
Eq. (6),
\begin{equation}
        C_0^{\overline{\rm MS}}(\alpha_s, \Lambda^2)
                = 1 + {\alpha_s\over \pi}
        \left(1-2{\Lambda^4\over Q^4}\right)
      + 1.67({\alpha_s\over \pi})^2
              \left(1-{\Lambda^4\over Q^4}
   \Big(2.89+1.50\ln{Q^2\over\Lambda^2}\Big)\right)
 + ...
\end{equation}
The above series is free of the IR renormalon
at $b=8\pi/\beta_0$ on the Borel plane.

To obtain the non-perturbative matrix
element of $F^2(\Lambda^2)$ (gluon condensate)
in dimensional regularization, we first write,
\begin{equation}
\langle 0|F^2(\Lambda^2)|0\rangle
 =\langle 0|F^2_R|0\rangle -
   \langle 0|F^2_R|0\rangle_{\rm pert}
   + \langle 0|F^2(\Lambda^2)|0\rangle_{\rm pert} \ .
\label{dim}
\end{equation}
According to the Joglekar-Lee theorems\cite{JL}, the matrix elements
of equations-of-motion operators and BRST-exact operators
vanish. Thus, the first
two matrix elements on the right-hand side of Eq. (\ref{dim}) depend on
the bare operator $F^2$ only, and
they transform homogeneously under a scale transformation.
Furthermore, the study on the trace anomaly of the
QCD energy-momentum tensor indicates
$\beta(g)F^2$ is renormalization scheme
and scale independent \cite{COL,MAN}. Thus,
$
\langle 0|F^2_R|0\rangle^{\overline {\rm MS} }-
   \langle 0|F^2_R|0\rangle_{\rm pert}^{\overline {\rm MS}}
  = Z\left[\langle 0|F^2|0\rangle^{\rm LAT} -
   \langle 0|F^2|0\rangle_{\rm pert}^{\rm LAT}\right]
$
where $Z = \beta^{\rm LAT}/\beta^{\rm MS} $ depends on
lattice and $\overline{\rm MS}$ $\beta$-functions.
Factorizing $Z$ in Eq. (\ref{dim}), we find
\begin{equation}
\langle 0|F^2(\Lambda^2)|0\rangle^{\overline {\rm MS}}
 =Z\left[\langle 0|F^2|0\rangle^{\rm LAT} -
   \langle 0|F^2|0\rangle^{\rm LAT}_{\rm pert}
   + Z^{-1}\langle 0|F^2(\Lambda^2)|0\rangle_{\rm pert}^{\overline{\rm
MS}}\right]
\label{dim2}
\end{equation}
The above is a practical
definition of the gluon condensate
that are renormalon-free and consistent with the $\overline{\rm MS}$
evaluation of
the coefficient functions.

The first two terms in Eq. (\ref{dim2})
are what has been used as a lattice definition
of the gluon condensate in the literature
\cite{BAN}. However,
$\langle 0|F^2|0\rangle^{\rm LAT}_{\rm pert}$
is an ill-defined perturbation series
due to the IR renormalon at $b=8\pi/\beta_0$. The
perturbation series has been numerically
evaluated up to eight-loops
by Di Renzo et al. in the quenched approximation\cite{DR} (see also
\cite{ACFP},
\begin{equation}
 \langle 0|F^2|0\rangle^{\rm LAT}_{\rm pert} =
       {48\over a^4}\left(1+4.01{\alpha_s(a)\over \pi}
   +64.08\Big({\alpha_s(a)\over \pi}\Big)^2+
  1321.86\Big({\alpha_s(a)\over \pi}\Big)^3 + ...\right)
\end{equation}
To extract the gluon condensate with a good accuracy,
one is faced with two opposite requirements for the lattice spacing $a$.
To ensure that the calculations are at the continuum limit,
one shall take as small $a$ as possible. On the other hand,
for small $a$, the cancellation of quartic divergences
requires extremely good accuracy in the perturbation series. Even
for $a=0.17$ fm ($\beta=5.7$), one needs to evaluate the
perturbation series to at least one percent. High-accuracy is
not possible without regularizing the IR renormalon in the series.
[Although for expansions with the bare lattice coupling, the smallest
term is postponed to higher orders, the intial decrease of the terms
is very slow due to large tadpole contributions.]
The last term in Eq. (\ref{dim2}) is introduced precisely for the
IR renormalon regularization. Using the one-loop relation between the lattice
and $\overline{\rm MS}$ couplings\cite{HH}, I find,
\begin{equation}
Z^{-1}\langle 0|F^2(\Lambda^2)|0\rangle_{\rm pert}^{\overline{\rm MS}}
  = {3\Lambda^4\over \pi^2}\left(1
       + \Big(20.87+1.25\ln{Q^2\over \Lambda^2} + 2.75\ln(Q^2a^2)\Big){
       \alpha_s(a)\over \pi} + ...\right)
\end{equation}
To determined the subtraction up to three-loops, we need
two-loop relations between lattice and
$\overline{\rm MS}$ couplings and the matrix element
$\langle 0|F^2(\Lambda^2)|0\rangle^{\overline{\rm MS}}
_{\rm pert}$ up to three-loops, the former
has recently been calculated by Luscher and
P. Weisz \cite{LU}. We wish to make a three-loop
subtraction calculation in the near future.

To summarize, I have introduced a minimal
subtraction scheme to regularize power divergences
of composite operators. When these operators are used
in Wilson's OPE, the coefficient functions
are automatically free of the IR renormalon singularities.
The matrix elements of the composite operators
can be calculated in non-perturbative methods
such as lattice QCD. Thus, within the scheme
one can meaningfully discuss power corrections
to desired accuracy. The scheme can be
straightforwardly applied to deep-inelastic sum rules where
the mixing of higher-twist operators
with lower-twist ones is an important issue\cite{JI0}. It can
also be used in the heavy quark effective
theory to define the pole masse for the heavy-quark
expansion\cite{BB}. This will be discussed in a separate
publication.

\acknowledgments
I wish to thank N. Christ, A. Mueller, J. Negele, U. Wiese
for useful conversations about the subject.

\end{document}